\documentclass[aps,pra,twocolumn]{revtex4-1}

\usepackage{times,mathptmx}

\usepackage{graphicx,pstricks,subfigure}

\usepackage{amsmath,amssymb}

\begin{document}

\title{
        Integral representation for scattering phase shifts 
        via the phase--amplitude approach
       }

\date{\today}

\author{D. Shu}
\author{I. Simbotin}
\author{R. C\^ot\'e}

\affiliation{Department of Physics, University of Connecticut,
   2152 Hillside Rd., Storrs, CT 06269-3046, USA}

\begin{abstract}

A novel integral representation for scattering phase shifts is
obtained based on a modified version of Milne's phase--amplitude
approach [W.~E.~Milne, Phys.~Rev.~\textbf{35}, 863 (1930)].  We
replace Milne's nonlinear differential equation for the amplitude
function $y$ with an equivalent linear equation for the envelope
$\rho=y^2$, which renders the integral representations highly amenable
to numerical implementations.  The phase shift is obtained directly
from Milne's phase function, which in turn is expressed in terms of
the envelope function.  The integral representation presented in this
work is fully general and it can be used for any type of scattering
potential, including the Coulomb potential.

\end{abstract}

\pacs{xx.yy.zz, aa.bb.cc, etc.etc.etc.....}


    \maketitle

\section{Introduction}

The phase--amplitude approach for Schr\"odinger's radial (or
one-dimensional) equation was pioneered by Milne \cite{milne}, and 
has since been used extensively in atomic and molecular physics
\cite{phase_amp_John_PRA49,John_and_Paul,robicheaux,second_order_WKB_wf_correction,
  Greene_John_PRL81,Chris_Greene_QDT_surface,Chris_Greene_QDT_milne_formalism,
crubellier,fabrikant-2012,olivier-raoult-2013,raoult-1988-prl,raoult-1994,raoult-2007},
in chemical physics
\cite{Multiple_well_Soo_Yin,electron_density_Jian_Min,quantum_chem_Light},
and in other areas of physics
\cite{plasma-rad,prb-2007,non_linear_optics_1,berry_phase_d_dimension,Pedrosa_PRA,BEC_phase,
  BEC_cosmology,ekpyrotic,TOV_neutron_star,Relic_Gravitons,Bilayer_graphene}.
Although it was originally intended for tackling bound states, the
phase--amplitude method is also applicable for scattering problems;
indeed, Milne's approach is especially suitable in the framework of
many-channel quantum defect theory
\cite{Raoult_AND_Mies,MQDT_Feshbach_osseni,Greene_fig,MQDT_Julienne}
because it makes it possible to construct optimal reference functions
in each scattering channel.

The virtues of Milne's approach stem from the fact that phases and
amplitudes are quantities which are well behaved in the energy domain,
even across channel thresholds.  Moreover, the phase--amplitude method
allows for highly efficient numerical implementations, because the
direct computation of highly oscillatory wave functions can be avoided
entirely; instead, any solution of the radial equation is 
evaluated accurately in terms of the amplitude and phase functions,
which have a simple radial dependence.

In this work we use a modified version of Milne's method to derive
a novel integral representation which yields the true value of the
scattering phase shift, despite the modulo $\pi$ ambiguity inherent in
its usual definition.  We begin with a simple integral representation
in Sec.~\ref{sec:delta}, which we generalize for the Coulomb case in
Sec.~\ref{sec:Coul}\@.  We give a brief discussion of envelope and
phase functions in Sec.~\ref{sec:rho-theta}, with additional details
of the phase--amplitude approach presented in the Appendix.  In
Sec.~\ref{sec:two-rho} we derive a two-envelope formula for phase
shifts, which is essential for high partial waves.  The computational
approach is discussed in Sec.~\ref{sec:asy}, and we illustrate the
practical usefulness of our integral representations with numerical
applications in Sec.~\ref{sec:results}\@.  Finally, we give concluding
remarks in Sec.~\ref{sec:concl}.

\section{Theory}
\label{sec:theory}

We consider the radial Schr\"odinger equation for a spherically
symmetric potential $V(R)$,
\begin{equation}\label{eq:Schroedinger}
\psi'' = U\psi,    \qquad    U=2\mu \big(V_{\rm eff}-E\big),
\end{equation}
where $V_{\rm eff}(R)=V(R)+\frac{\ell(\ell+1)}{2\mu R^2}$ is the effective
potential,  $\mu$ is the reduced mass of the two particles undergoing
scattering, and  $E>0$ is the energy in the center-of-mass frame.  Atomic
units are used throughout.

Assuming that the interaction potential $V(R)$ vanishes faster than
$R^{-1}$ asymptotically, the physical solution $\psi(R)$ has the well
known asymptotic behavior
\begin{equation}\label{eq:psi-asy-delta}
\psi(R) \xrightarrow{R\to\infty} \sin(kR-\ell\textstyle\frac\pi2+\delta_\ell),
\end{equation}
which yields the scattering phase shift $\delta_\ell$ for partial wave
$\ell$.  In the equation above, $k=\sqrt{2\mu E}$ is the momentum for
the relative motion.  The phase shift is usually obtained from
the matching condition,
\begin{equation}\label{eq:psi-match}
\psi(R) = A f(R) + B g(R),
\end{equation}
where $f$ and $g$ are exact solutions of the radial equation,
specified by their asymptotic behavior,
\begin{eqnarray}
f(R)  &   \xrightarrow{R\to\infty} &   \sin(kR-\ell\textstyle\frac\pi2)
\label{eq:f-asy}
\\
g(R)  &    \xrightarrow{R\to\infty} &  \cos(kR-\ell\textstyle\frac\pi2).
\label{eq:g-asy}
\end{eqnarray}
From the equations above we have $\delta_\ell=\arctan(B/A)$, which
yields the phase shift modulo $\pi$, i.e.,
$\delta_\ell\in[-\frac\pi2,\;\frac\pi2]$.  To simplify our notation,
we shall omit the label $\ell$ for $\psi$, $f$, $g$ and other related
quantities, except for the phase shift $\delta_\ell$.

    \subsection{Integral representation for the \emph{full} phase shift}
                  \label{sec:delta}

We now derive an expression for $\delta_\ell$ which does not rely on
the explicit evaluation of wave functions; instead, the phase shift
will be extracted from an $R$-dependent phase function.  Our approach
follows Milne's phase--amplitude method~\cite{milne}, which we
review in appendices A, B, C\@.  We emphasize that the \emph{true}
value of $\delta_\ell$ will be obtained, despite the modulo $\pi$
ambiguity inherent in its customary definition.

We first introduce the envelope function
\begin{equation}\label{eq:rho=f2+g2}
\rho = f^2 + g^2,
\end{equation}
with $f$ and $g$  exact solutions of the radial
equation~(\ref{eq:Schroedinger}) obeying the asymptotic
behavior~(4,~5).  The phase function $\theta$ is constructed by
integrating
\begin{equation}\label{eq:theta-prime}
\theta' \equiv \frac{d\theta}{dR} = \frac k\rho.
\end{equation}
We remark that $\theta(R)$ is defined up to an integration constant,
which can be chosen freely;  a judicious choice guided by the
computational strategy shall be made in Sec.~\ref{sec:rho-theta}\@.

As shown by Milne~\cite{milne}, the general solution of the radial
equation~(\ref{eq:Schroedinger}) can be represented exactly in terms
of $\rho$ and $\theta$.  In particular,
the physical solution reads
\begin{equation}\label{eq:milne-psi}
\psi(R) = \sqrt{\rho(R)} \sin[\theta(R)-\theta(0)].
\end{equation}
Note that $\psi(R)$ vanishes explicitly at $R=0$, while
Eqs.~(\ref{eq:f-asy},~\ref{eq:g-asy}) ensure a very simple
asymptotic behavior for $\rho$ and $\theta$,
\begin{eqnarray}
\rho(R)    & \xrightarrow{R\to\infty}  &  1  
\label{eq:rho-to-infty=1}\\
\theta(R)  & \xrightarrow{R\to\infty}  &  kR +{\rm const.}
\label{eq:theta-to-infty=kR}
\end{eqnarray}
We now define the reduced phase
\begin{equation}\label{eq:theta-tilde-def}
\tilde\theta(R)  \equiv  \theta(R) -kR,
\end{equation}
and use Eqs.~(\ref{eq:rho-to-infty=1},~\ref{eq:theta-to-infty=kR}) to
find the asymptotic behavior of Milne's
parametrization~(\ref{eq:milne-psi}),
\[
\psi(R)  \xrightarrow{R\to\infty} \sin[kR+\tilde\theta(\infty)-\theta(0)],
\]
which is identical to the asymptotic behavior of $\psi$ in
Eq.~(\ref{eq:psi-asy-delta}).  Consequently, we obtain
\begin{equation}\label{eq:delta=theta-general}
\delta_\ell -\ell\textstyle\frac\pi2 = \tilde\theta(\infty)-\theta(0).
\end{equation}
Making use of Eq.~(\ref{eq:theta-tilde-def}) we have
$\theta(0)=\tilde\theta(0)$, and
Eqs.~(\ref{eq:theta-prime},~\ref{eq:theta-tilde-def}) yield
$\tilde\theta'=\frac k\rho-k$.  Hence,
Eq.~(\ref{eq:delta=theta-general}) can be recast as an integral
representation,
\begin{eqnarray}
\delta_\ell -\ell{\textstyle\frac\pi2}
 &=&  \tilde\theta(\infty)-\tilde\theta(0)
\nonumber
\\ 
 &=&  k\!\int_0^\infty \!\! dr \bigg[\frac 1 {\rho(r)} -1\bigg].\label{eq:Delta=int}
\end{eqnarray}
We remark that the reduced phase $\tilde\theta(R)$ defined in
Eq.~(\ref{eq:theta-tilde-def}) cannot be regarded as the nontrivial
phase contribution, as it also includes the Bessel contribution
(due to the centrifugal term).  Therefore, strictly speaking, the
equation above yields the \emph{full} phase shift
($\delta_\ell-\ell\frac\pi2$).

\subsection{Generalization to potentials with a  Coulomb term}

     \label{sec:Coul}

So far, we assumed that the potential $V(R)$ vanishes faster than
$R^{-1}$ asymptotically.  In this section we consider the general case
when $V(R)$ contains a Coulomb term,
\begin{equation}\label{eq:Vcoul}
V_C(R) = \frac{Z_1Z_2} R.
\end{equation}
with $Z_{1,2}$ the electric charges of the two colliding particles.
The remainder ($V-V_C$) of the interaction potential is responsible
for the phase shift $\delta_\ell$, which is obtained from the well
known asymptotic behavior
\begin{equation}\label{eq:psi-asy-Coul}
\psi(R)  \xrightarrow{R\to\infty} \sin\big[kR-\textstyle\frac C k \ln(2kR)
          +\eta_\ell -\ell\frac\pi2 +\delta_\ell \big],
\end{equation}
where $\eta_\ell=\arg\Gamma(\ell+1 + i\frac C k)$ is the Coulomb phase
shift~\cite{seaton-2002}, and $C\equiv\mu Z_1Z_2$.
Following the same steps as in the previous section, we use again
Milne's parametrization~(\ref{eq:milne-psi}) for the physical wave
function, namely $\psi(R)=\sqrt{\rho(R)}\sin[\theta(R)-\theta(0)]$,
with the phase $\theta$ behaving asymptotically as
\begin{equation}\label{eq:theta-to-infty-C}
\theta(R)   \xrightarrow{R\to\infty} 
            kR-\textstyle\frac C k \ln(2kR) +{\rm const},
\end{equation}
which is the generalized version of Eq.~(\ref{eq:theta-to-infty=kR}).
Accordingly, the reduced phase is again defined such that it is finite
asymptotically,
\begin{equation}\label{eq:theta-tilde-C}
\tilde\theta(R) \equiv \theta(R) -kR+\textstyle\frac C k \ln(2kR),
\end{equation}
and thus the full phase shift reads
\begin{equation}
\label{eq:delta=theta-general-C}
\delta_\ell - \ell\textstyle\frac\pi2 + \eta_\ell
 = \tilde\theta(\infty)-\theta(0),
\end{equation}
which is the general form of Eq.~(\ref{eq:delta=theta-general}).

In the Coulomb case, an integral representation can only be written if
we divide the radial domain in two intervals.  Indeed, unlike the
previous section, $\tilde\theta$ now diverges logarithmically when
$R\to0$.  Thus, we shall employ $\theta(R)$ for $R\leq R_0$ and
$\tilde\theta(R)$ for $R\geq R_0$, with $R_0$ fixed arbitrarily.
Specifically, we have
\begin{eqnarray*}
\theta(R_0)-\theta(0) &=& k\!\int_0^{R_0}\!\!\frac{dr}{\rho(r)}
\\
\tilde\theta(\infty)-\tilde\theta(R_0) &=&
k\!\int_{R_0}^\infty\!\!dr\left[\frac1{\rho(r)}-1+\frac{C}{k^2r}\right].
\end{eqnarray*}
Adding the two equations above, and  making use of
Eq.~(\ref{eq:theta-tilde-C}) at $R=R_0$, we find
\begin{eqnarray}
\delta_\ell - \ell\textstyle\frac\pi2 + \eta_\ell
 &=& \tilde\theta(\infty)-\theta(0)
\nonumber\\
 &=& k\!\!\int_0^{R_0}\!\!dr\left[\frac{1}{\rho(r)}-1\right] + {\textstyle\frac C k} \ln(2kR_0)
\nonumber\\
 &+& k\!\!\int_{R_0}^\infty\!\!dr\left[\frac1{\rho(r)}-1+\frac{C}{k^2r}\right],
\label{eq:Delta=int-C}
\end{eqnarray}
which represents the generalization of Eq.~(\ref{eq:Delta=int}).
Indeed, in the absence of the Coulomb term, i.e., setting $C=0$ in the
equations above, we have $\eta_\ell=0$, and we recover the results of
Sec.~\ref{sec:delta}\@.  We emphasize that the expression in
Eq.~(\ref{eq:Delta=int-C}) is independent of $R_0$, which we
illustrate with numerical results in Sec.~\ref{sec:results-Coulomb}\@.
Finally, we remark that
Eqs.~(\ref{eq:Delta=int},~\ref{eq:Delta=int-C}) yield the true value
of $\delta_\ell$ unambiguously (not modulo $\pi$), and in the case of
a purely Coulombic potential, Eq.~(\ref{eq:Delta=int-C}) yields the
true value of $\eta_\ell$.

\subsection{Envelope and phase functions}
              \label{sec:rho-theta}

In order to use the approach outlined above in numerical applications,
it is necessary to devise a reliable method for computing the envelope
directly, rather than using Eq.~(\ref{eq:rho=f2+g2}).  As shown in
App.~\ref{app:rho-lin-eq}, $\rho$ obeys a linear differential
equation,
\begin{equation}  \label{eq:rho-linear}
\rho''' - 4U\rho' - 2 U'\rho = 0.
\end{equation}
Therefore, we now regard the envelope $\rho=f^2+g^2$ in
Eq.~(\ref{eq:rho=f2+g2}) as a particular solution of
Eq.~(\ref{eq:rho-linear}).  Namely, we impose the asymptotic boundary
condition $\rho(R)\to1$, which makes the solution unique.  As we shall
see in Sec.~\ref{sec:asy}, we initialize the envelope at $R=\infty$
and we propagate it inward; accordingly, we also propagate the reduced
phase $\tilde\theta$ inward from $R=\infty$.

Recall that Eq.~(\ref{eq:theta-prime}) allows for an 
integration constant to be chosen freely when constructing the phase
$\theta(R)$ or $\tilde\theta(R)$.  The integration constant can be
fixed, e.g., by setting the value of $\theta(0)$, or the value of
$\tilde\theta(\infty)$.  We prefer the latter, which is suitable when
employing the inward propagation mentioned above; specifically, we
choose
\[
\tilde\theta(\infty) =0,
\]
and thus Eq.~(\ref{eq:delta=theta-general}) reads
\begin{equation}\label{eq:Delta=theta-at-zero}
\delta_\ell = \ell\textstyle\frac\pi2 -\theta(0).
\end{equation}

The reduced phase is constructed by direct integration; in the absence
of a Coulomb interaction term, we have
\begin{equation}\label{eq:theta-tilde=int}
 \tilde\theta(R)
 = k\!\int_R^\infty\!\!\!dr\,\frac{\tilde\rho(r)}{\rho(r)},
\end{equation}
where $\tilde\rho$ denotes the reduced envelope
\[
\tilde\rho \equiv \rho -1.
\]
In the Coulomb case we make use of
$\tilde\theta'(r)=\frac{k}{\rho(r)}-k+\frac{C}{kr}$, see
Eqs.~(\ref{eq:theta-prime},~\ref{eq:theta-tilde-C}), and thus the
reduced phase reads
\begin{equation}\label{eq:theta-tilde=int-C-rho}
 \tilde\theta(R)
 = k\!\int_R^\infty\!\!\!dr\,\left[1-\frac1{\rho(r)}-\frac{C}{k^2r}\right].
\end{equation}
In order to show that the integral above is well defined and yields a
reduced phase obeying $\tilde\theta(\infty)=0$, and also to justify
that the full phase $\theta(R)$ has the asymptotic
behavior~(\ref{eq:theta-to-infty-C}), we write the envelope as an
asymptotic series,
\begin{equation}\label{eq:rho-ansatz}
\rho(R) = 1 +\sum_{n\geq1}\frac{b_n}{R^n}.
\end{equation}
Assuming the potential has the long-range behavior
\[
V(R) = \sum_{n\geq1}\frac{C_n}{R^n}, \qquad {\rm with}\ \ C_1=C=\mu Z_1Z_2,
\]
we substitute the ansatz~(\ref{eq:rho-ansatz}) in
Eq.~(\ref{eq:rho-linear}), and we obtain the coefficients $b_n$.  In
particular, for $n=1$ we have $b_1=C/k^2$, and thus the asymptotic
behavior of the envelope reads
\[
\rho(R) \approx 1+\frac C{k^2R} +\frac{b_2}{R^2} + \cdots.
\]
Substituting the result above in Eq.~(\ref{eq:theta-prime}) yields
$\theta'\approx k-\frac C{kR}+\mathcal O(\frac1{R^2})$, which upon
integration confirms Eq.~(\ref{eq:theta-to-infty-C}), while the
reduced phase in Eq.~(\ref{eq:theta-tilde=int-C-rho}) has the
asymptotic behavior
\[
\tilde\theta(R) \approx \frac{\tilde b_1}R + \frac{\tilde b_2}{R^2}
 + \cdots \xrightarrow{R\to\infty}0,
\]
where the coefficients $\tilde b_n$ are expressed in terms of $b_n$,
e.g., $\tilde b_1=b_2-b_1^2$.

For computational purposes, it is advantageous to employ the reduced
envelope
\begin{equation}
\label{eq:rho-Coul}
\tilde\rho\equiv  \rho -1 -\textstyle\frac C {k^2r}.
\end{equation}
Thus, we recast Eq.~(\ref{eq:theta-tilde=int-C-rho}) in terms of
$\tilde\rho$,
\begin{equation}\label{eq:theta-tilde=int-C-rho-tilde}
\tilde\theta(R) = k\!\!\int_R^\infty\!\!dr \frac 1{\rho(r)}
\left[\left(1-\textstyle\frac C{k^2r}\right)\tilde\rho(r)
-\left(\textstyle\frac C{k^2r}\right)^2\right].
\end{equation}
The full phase shift in Eq.~(\ref{eq:Delta=int-C}) can now be expressed
in a form suitable for computation,
\begin{eqnarray}
\delta_\ell &-& \ell{\textstyle\frac\pi2} + \eta_\ell
 = -\theta(0)\nonumber
\\
 &=& k\!\!\int_0^{R_0}\!\!\frac{dr}{\rho(r)}
 -kR_0 + {\textstyle\frac C k} \ln(2kR_0) -\tilde\theta(R_0).
\label{eq:Delta-C}
\end{eqnarray}

We emphasize that $\tilde\theta$ should be computed using
Eq.~(\ref{eq:theta-tilde=int-C-rho-tilde}), because the integrand in
Eq.~(\ref{eq:theta-tilde=int-C-rho}) suffers from cancellation at
large $r$.  Thus, in the asymptotic region, $\tilde\rho$ should be
obtained directly, rather than $\rho$ itself; namely, we substitute
$\rho=\tilde\rho+1+C/k^2R$ in Eq.~(\ref{eq:rho-linear}) which becomes
an equation for $\tilde\rho$.  The numerical approach for solving the
envelope equation is described in Sec.~\ref{sec:asy}, where we show
that the entire asymptotic region can be treated in a numerically
exact fashion by mapping it onto a finite interval and using a
spectral Chebyshev method.

    \subsection{Two-envelope formula for phase shifts}
                 \label{sec:two-rho}

We now derive an integral representation which makes it possible to
compute the phase shift $\delta_\ell$ directly.  Recall that the
simple integral representations (\ref{eq:Delta=int}) and
(\ref{eq:Delta=int-C}) yield the \emph{full} phase shift (including
Bessel and Coulomb contributions).  As shown in
Sec.~\ref{sec:results-C3}, accurate values of $\delta_\ell$ at high
$\ell$ cannot be obtained using Eq.~(\ref{eq:Delta=int}).  Thus, in
this section we formulate a two-envelope approach which can be used to
compute accurate phase shifts for all partial waves.  For the sake of
generality, we consider two different potentials $V_1$ and $V_2$, each
containing the same Coulomb interaction term (if present).  For a
given scattering energy, $E>0$, and a partial wave $\ell$, we make use
of Eq.~(\ref{eq:delta=theta-general-C}) for each potential, and we
employ the convenient choice
$\tilde\theta_1(\infty)=\tilde\theta_2(\infty)$ to find
\begin{equation}\label{eq:delta2-1}
\delta_\ell^{(2)} - \delta_\ell^{(1)} = \theta_1(0) - \theta_2(0),
\end{equation}
Using $\theta'_{1,\,2}=k/\rho_{1,\,2}$, see
Eq.~(\ref{eq:theta-prime}), the phase difference above can be recast
as an integral,
\begin{equation}\label{eq:delta=rho2-1}
\delta_\ell^{(2)} -\delta_\ell^{(1)}
 = k\!\int_0^\infty\!\!dr\Big[\frac1{\rho_2(r)}-\frac1{\rho_1(r)}\Big].
\end{equation}
Although both $\theta_{1,\,2}(R)$ diverge when $R\to\infty$, the
integral above is finite because the phase difference
$\theta_1(R)-\theta_2(R)=\tilde\theta_1(R)-\tilde\theta_2(R)$ vanishes
asymptotically.  Note that in the asymptotic region we have
$\rho_1\approx\rho_2$, which can also hold in the inner region if
$V_1\approx V_2$.  Thus, Eq.~(\ref{eq:delta=rho2-1}) will suffer from
catastrophic cancellation,  rendering it unsuitable for
numerical applications.  Nevertheless, we show next that  a
computationally robust integral representation based on the
two-envelope approach can be formulated.

We choose $V_1\equiv V_{\rm ref}$ as a reference potential (with the
corresponding effective potential including both the centrifugal and
Coulomb terms, see below), while $V_2\equiv V=V_{\rm ref}+\hat V$ is the
full interaction potential.  The reduced envelope and phase are now
defined relative to the corresponding reference quantities,
\begin{eqnarray}
\hat\rho  &=&  \rho-\rho_{\rm ref}  \nonumber
\\
\hat\theta &=& \theta-\theta_{\rm ref}.
\label{eq:theta-hat}
\end{eqnarray}
We  employ a nontrivial reference problem by setting
\begin{equation}\label{eq:Uref}
U_{\rm ref}(R) = -k^2 + \frac{\ell(\ell+1)} {R^2} +\frac{2C}R,
\qquad C=\mu Z_1Z_2, 
\end{equation}
and we use of Eqs.~(\ref{eq:delta2-1},~\ref{eq:delta=rho2-1}) with
$U_1=U_{\rm ref}$ given above and $U_2=U=U_{\rm ref}+2\mu\hat V$\@.
Thus, the Bessel phase shift ($-\ell\frac\pi2$) and the Coulomb phase
shift $\eta_\ell$ are both eliminated, and the phase shift
$\delta_\ell$ reads
\begin{eqnarray}
\delta_\ell
 &=& -\hat\theta(0)\nonumber\\
 &=& -k\!\int_0^\infty\!\!dr\frac{\hat\rho(r)}{\rho(r)\rho_{\rm ref}(r)}.
 \label{eq:delta=theta-hat-0}
\end{eqnarray}
The reference envelope is the solution of Eq.~(\ref{eq:rho-linear}),
\begin{equation}\label{eq:rho-ref}
\rho_{\rm ref}''' - 4U_{\rm ref}\,\rho'_{\rm ref} - 2 U_{\rm ref}'\,\rho_{\rm ref}
 = 0,
\end{equation}
with $U_{\rm ref}(R)$ given in Eq.~(\ref{eq:Uref}), while the reduced
envelope obeys a non-homogeneous differential equation,
\begin{equation}\label{eq:rho-hat-nonhmg}
\hat\rho''' - 4U\hat\rho' - 2U'\hat\rho = 4\hat U\rho'_{\rm ref}+2\hat
U'\rho_{\rm ref},
\end{equation}
which was obtained by combining Eqs.~(\ref{eq:rho-linear})
and~(\ref{eq:rho-ref}).  In the equation above we used the notation
$\hat U\equiv U-U_{\rm ref}=2\mu\hat V$.
Note that in the absence of a Coulomb term, we use $U_{\rm ref}=
-k^2+\frac{\ell(\ell+1)}{R^2}$, with $V_{\rm ref}=0$ and $\hat V=V$,
which is illustrated with numerical results in
Sec.~\ref{sec:results-C3}\@.  Finally, we remark that in the absence
of a Coulomb term one can also use the trivial choice $U_{\rm
  ref}=-k^2$ with $\theta_{\rm ref}(R)=kR$, which yields $\hat\theta$
identical to $\tilde\theta$ in Eq.~(\ref{eq:theta-tilde-def}), thus
recovering the integral representation of the full phase shift given
in Sec.~\ref{sec:delta}\@.

In practical applications, one first solves Eq.~(\ref{eq:rho-ref}) for
the reference envelope, which is subsequently used in
Eq.~(\ref{eq:rho-hat-nonhmg}).  The latter is solved to obtain
$\hat\rho$, and thus the full envelope is obtained: $\rho=\rho_{\rm
  ref}+\hat\rho$.  We remark that the numerical approach used for the
homogeneous envelope equation, see Sec.~\ref{sec:asy}, can also be
employed for the non-homogeneous differential
equation~(\ref{eq:rho-hat-nonhmg}).  The two-envelope approach is
fully general, but is especially useful when $\hat{U}=2\mu\hat V$ is
small, such that $|\hat\rho|\ll\rho\approx\rho_{\rm ref}$, and thus
$\hat\theta$ and $\delta_\ell$ will also be small.

\section{Computational approach}

          \label{sec:asy}

The integral representations derived in this article are valid in
general; however, they are useful in numerical applications only if
the integrands are well behaved.  Specifically, the envelope should
behave in a non-oscillatory fashion, which in turn ensures the
smoothness of the phase function.  In practice, there is considerable
difficulty in finding the smooth envelope
\cite{robicheaux,sidky_phys_essay}, because the general solution of
the envelope equation has an oscillatory behavior, as mentioned in
App.~\ref{app:milne}\@.  In this section, a novel computational
strategy for obtaining the \emph{smooth} envelope in the asymptotic
region is presented.  We emphasize that finding the unique, smooth
solution is critically important for the efficiency and accuracy of
numerical schemes using our integral representations, and for using
the phase--amplitude method in general.

For clarity, we assume that $V(R)$ vanishes faster than $R^{-1}$
asymptotically.  As discussed in Sec.~\ref{sec:delta} (see also
App.~\ref{app:q-invariant}), we employ the initial condition $\rho=1$
at $R=\infty$ when solving Eq.~(\ref{eq:rho-linear}).  However, rather
than using the radial variable, it is highly advantageous to
reformulate the envelope equation by \emph{mapping} the asymptotic
radial domain onto a finite interval.  We developed a convenient and
efficient numerical implementation based on a simple change of
variable,
\[
x=\frac1 R,
\]
which allows to take fully into account the long-range tail of any
potential.  Thus, the \emph{infinite} radial interval $R_1<R<\infty$ is now
mapped on a compact interval, $x_1>x>0$, with $x_1=\frac1{R_1}$.  The
boundary $R_1$, i.e., the size of the interval $[0,\;x_1]$, will be
chosen to ensure the desired level of accuracy.

We now regard the envelope as an $x$-dependent function, and we
present the computational approach for finding the smooth solution of
the envelope equation inside the interval $[0,\;x_1]$.  First, we
rewrite Eq.~(\ref{eq:rho-linear}) using $x=\frac 1 R$ as the
independent variable,
\begin{equation}\label{eq:rho-x}
 x^4\dddot\rho +6x^3\ddot\rho +6x^2\dot\rho -4U\dot\rho -2\dot U \rho = 0,
\end{equation}
where dots above symbols denote derivatives with respect to $x$,
e.g., $\dot\rho=d\rho/dx$.  Recall that $U=2\mu V_{\rm eff}-k^2$.
Next, we define
\[
u \equiv \dot\rho,
\]
and we regard it as the unknown function.  Making use of the
initial condition $\rho=1$ at $x=0$, we write
\[
\rho(x)=1+\!\int_0^x\!dt\,u(t),
\]
which we substitute in the last term of Eq.~(\ref{eq:rho-x}) to obtain
\begin{equation}\label{eq:Mu}
\Big( 4k^2 -8\mu V_{\rm eff}
    + x^4D_x^2 +6x^3D_x +6x^2 -4\mu\dot V_{\rm eff}S_x\Big)u
  = 4\mu\dot V_{\rm eff}.
\end{equation}
The operators $D_x$ and $S_x$ read
\[
D_xu=\dot u=\frac{du}{dx},  \qquad  S_xu=\!\!\int_0^x\!\!dt\,u(t).
\]

We solve Eq.~(\ref{eq:Mu}) using a spectral Chebyshev
method \cite{Clenshaw_Curtis,El-gendi,Greengard,George_IEM,Mihaila,George},
i.e., we employ a small number of Chebyshev polynomials $T_n(x)$ with
$n=0,1,2,\ldots,N-1$, which are mapped onto the interval $[0,x_1]$.
We expand the unknown function $u(x)$ in the Chebyshev basis, and the
operators $D_x$ and $S_x$ are represented as finite ($N\times N$)
matrices~\cite{Greengard,George_IEM}.  Thus, Eq.~(\ref{eq:Mu}) becomes a
simple linear system,
\[
\mathbf {M A = B},
\]
where the column $\mathbf A$  contains the  Chebyshev coefficients of our unknown function,
\[
u(x) = \sum_{n=0}^{N-1} A_n T_n(x),
\]
$\mathbf B$ contains the Chebyshev coefficients for the expansion
\[
4\mu\dot V_{\rm eff}(x) =  \sum_{n=0}^{N-1} B_n T_n(x),
\]
and $\mathbf M$ is the matrix of the operator in Eq.~(\ref{eq:Mu}),
\[
M=4k^2 -8\mu V_{\rm eff} + x^4D_x^2 +6x^3D_x +6x^2 -4\mu\dot V_{\rm eff}S_x.
\]
$M$ is singular, but its matrix ($\mathbf M$) in the small Chebyshev
basis is well conditioned and thus yields highly accurate solutions:
$\mathbf{A=M^{-1}B}$\@.  The smooth envelope is obtained as the unique
solution, because all the other solutions oscillate infinitely fast
near $x=0$ ($R\to\infty$) and they are thus eliminated; indeed, highly
oscillatory behavior cannot be accommodated by the small number of
Chebyshev polynomials.  We emphasize that the linearity of the
envelope equation is crucially important for the feasibility of the
approach presented here.  Finally, the solution obtained inside the
interval $[0,\;x_1]$ can now be used to initialize the propagation for
$x>x_1$, i.e., $R<R_1$.  A detailed description of our computational
approach will be published elsewhere.

\begin{figure}[t]
\includegraphics[width=\linewidth]{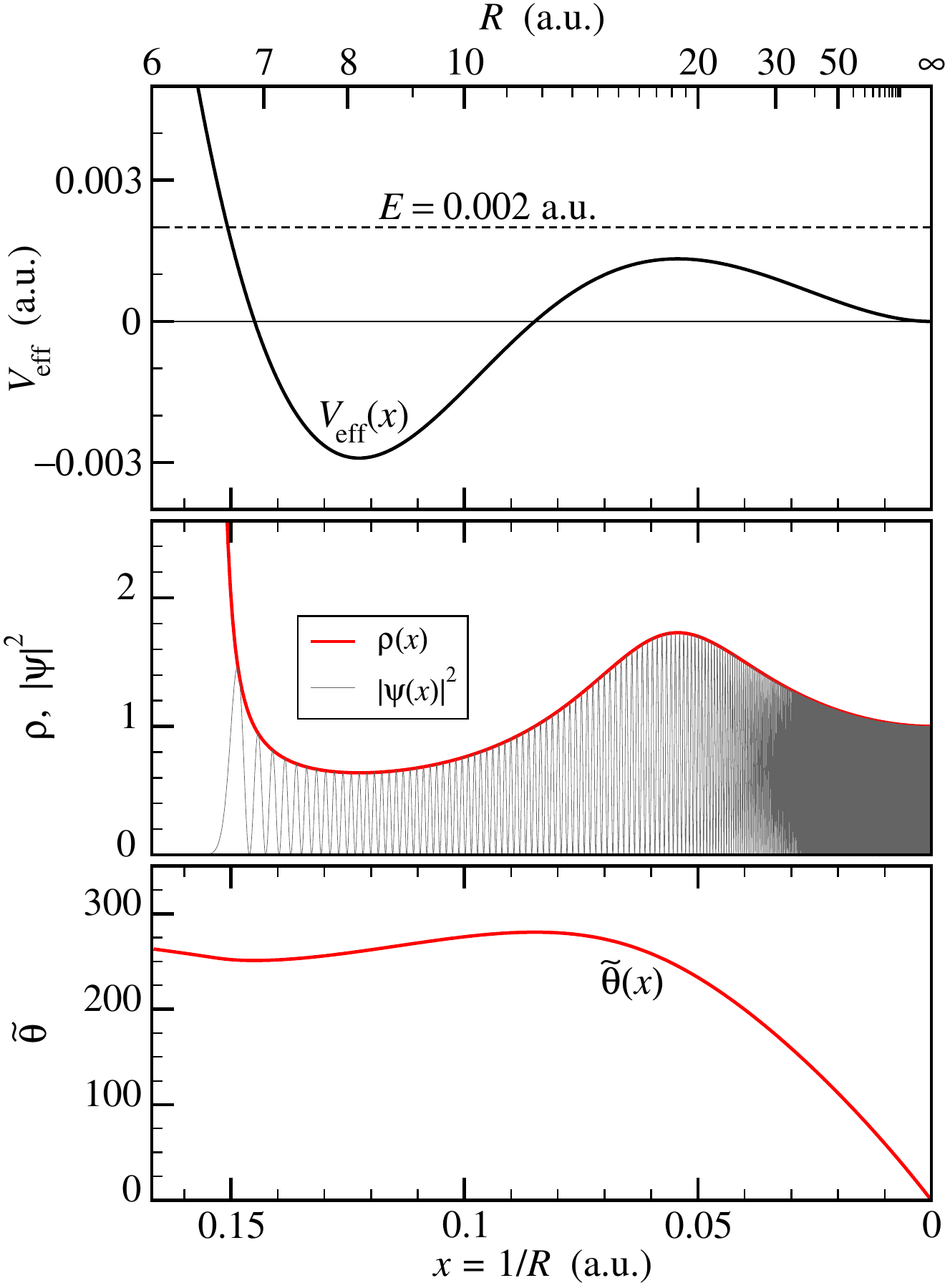}
\caption{\label{fig:x-rho-psi} Upper: effective potential (for
  $\ell=475$) as a function of $x=\frac1 R$\@.
  $V(R)$ is given in Eq.~(\ref{eq:vpotC3}).  The values of $R$ are indicated
  at the top.  The horizontal dashed line marks the energy $E$\@.
  Middle: thin  line (gray) for the wave function,  thick  line (red)
  for the envelope.  Lower: reduced phase $\tilde\theta(x)$.
 }
\end{figure}

The advantage of the phase--amplitude method combined with the change
of variable $x=\frac1 R$ is readily apparent in
Fig.~\ref{fig:x-rho-psi}, where we show the $x$-dependence of the
reduced phase $\tilde\theta(x)$ and the envelope $\rho(x)$ along with
$|\psi|^2$ for $\ell=475$ and $E=0.002$~a.u., for the potential energy
used in Sec.~\ref{sec:results-C3}, see Eq.~(\ref{eq:vpotC3}).  Note that the
wave function in Fig.~\ref{fig:x-rho-psi} was evaluated numerically
using Eq.~(\ref{eq:milne-psi}), i.e.,
$\psi=\sqrt\rho\sin(\theta-\theta_{R=0})$.

When the potential has a repulsive wall at short range, the
formulation of the phase--amplitude approach based on the change of
variable $x=\frac1 R$ can be employed for the entire radial domain, as
shown in Fig.~\ref{fig:x-rho-psi}.  Indeed, the repulsive wall makes
it possible to stop the inward propagation at $R_{\rm min}>0$, which
corresponds to a finite value $x_{\rm max}=\frac1{R_{\rm min}}$.  We
remark that, when the inward propagation of the reduced phase
$\tilde\theta$ approaches the repulsive wall, it is convenient to
convert it to the full phase $\theta$ using
Eq.~(\ref{eq:theta-tilde-def}) at a point $R_0$ just outside the inner
wall; for the remainder of the radial domain ($R<R_0$), one should
propagate $\theta(R)$ instead of $\tilde\theta(R)$, because the former
converges much faster than the latter.  Indeed, if $V_{\rm eff}\to+\infty$
when $R\to0$, we have $\rho\to\infty$ and
$\theta'\to0$, while $\tilde\theta'\to-k$.  In other words,
$\theta(0)$ should be computed as
\begin{equation}\label{eq:theta0-R0}
\theta(0) = kR_0
 +k\!\!\int_{R_0}^\infty\!\!dr\,\frac{\tilde\rho(r)}{\rho(r)}
 -k\!\!\int_{R_{\rm min}}^{R_0} \frac{dr}{\rho(r)},
\end{equation}
which is independent of $R_0$.  We remark that the integration need
not extend fully to $R=0$, because $R_{\rm min}$ is chosen inside the
repulsive wall to ensure the contribution of the interval $0<R<R_{\rm
  min}$ is entirely negligible.  Consequently, the radial domain can
be safely restricted to $R_{\rm min}<R<\infty$, which is mapped onto
the compact interval $x_{\rm max}>x>0$.

\section{Examples}
  \label{sec:results}

We now apply the integral representations and show that they yield
highly accurate results.
Our first example is the Coulomb potential, which we use as a test
case for the integral representation~(\ref{eq:Delta-C}).  The
two-envelope formula, see Eqs.~(\ref{eq:delta2-1})
and~(\ref{eq:delta=theta-hat-0}), will be employed and tested in
Sec.~\ref{sec:results-C3}.

\subsection{The Coulomb potential}
\label{sec:results-Coulomb}

\begin{figure}[t]
\includegraphics[width=0.99\linewidth]{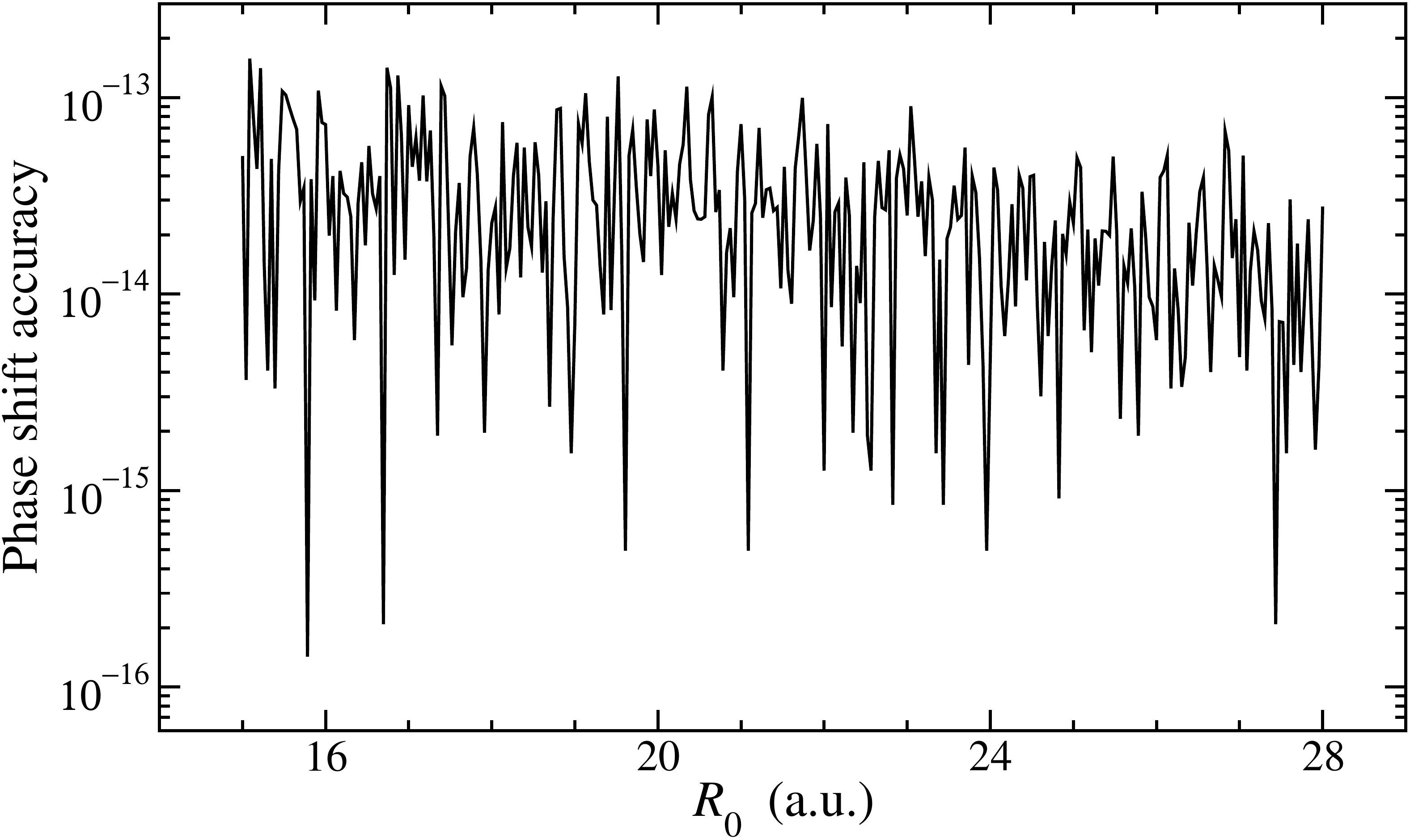}
\caption{\label{fig:eta-error} Relative error of the computed Coulomb
  phase shift $\eta_\ell$ for $C=-1$, $k=0.1$ and $\ell=5$.
  Equation~(\ref{eq:eta}) yields a highly accurate result that is
  independent of $R_0$.}
\end{figure}

In the case of a purely Coulombic potential, Eq.~(\ref{eq:Delta-C})
yields the Coulomb phase shift,
\begin{equation}\label{eq:eta}
\eta_\ell =\ell\textstyle\frac\pi2 -kR_0 + \frac C k \ln(2kR_0) \displaystyle
 +k\!\!\int_0^{R_0} \!\!\frac{dr}{\rho(r)} -\tilde\theta(R_0),
\end{equation}
with $\tilde\theta$ given Eq.~(\ref{eq:theta-tilde=int-C-rho-tilde}).
The result above is independent of $R_0$, as depicted in
Fig.~\ref{fig:eta-error}.  Indeed, we show that our approach is robust
and accurate by comparing the value of $\eta_\ell$ obtained using
Eq.~(\ref{eq:eta}) with the exact value
$\eta^\Gamma_\ell\equiv\arg\Gamma(\ell+1+i\frac C k)$ for $C=-1$,
$k=0.1$ and $\ell=5$.  Our integral representation yields the value
$\eta_\ell=-20.22421961527$, while the analytical expression gives its
value modulo 2$\pi$, namely $\eta^\Gamma_\ell=-1.3746636937335435$.
Their difference equals an integer multiple of $2\pi$ to a high degree
of precision: $(\eta_\ell-\eta^\Gamma_\ell)/2\pi=-3(1\pm10^{-13})$.
Figure~\ref{fig:eta-error} depicts the relative error
$|(\eta_\ell-\eta^\Gamma_\ell+6\pi)/\eta_\ell|$ as a function of
$R_0$.

\subsection{Results for an inter-atomic potential with long-range
  behavior of the type $V(R)\sim-\frac{C_3}{R^3}$}

              \label{sec:results-C3}

\begin{figure}[b]
\includegraphics[width=\linewidth]{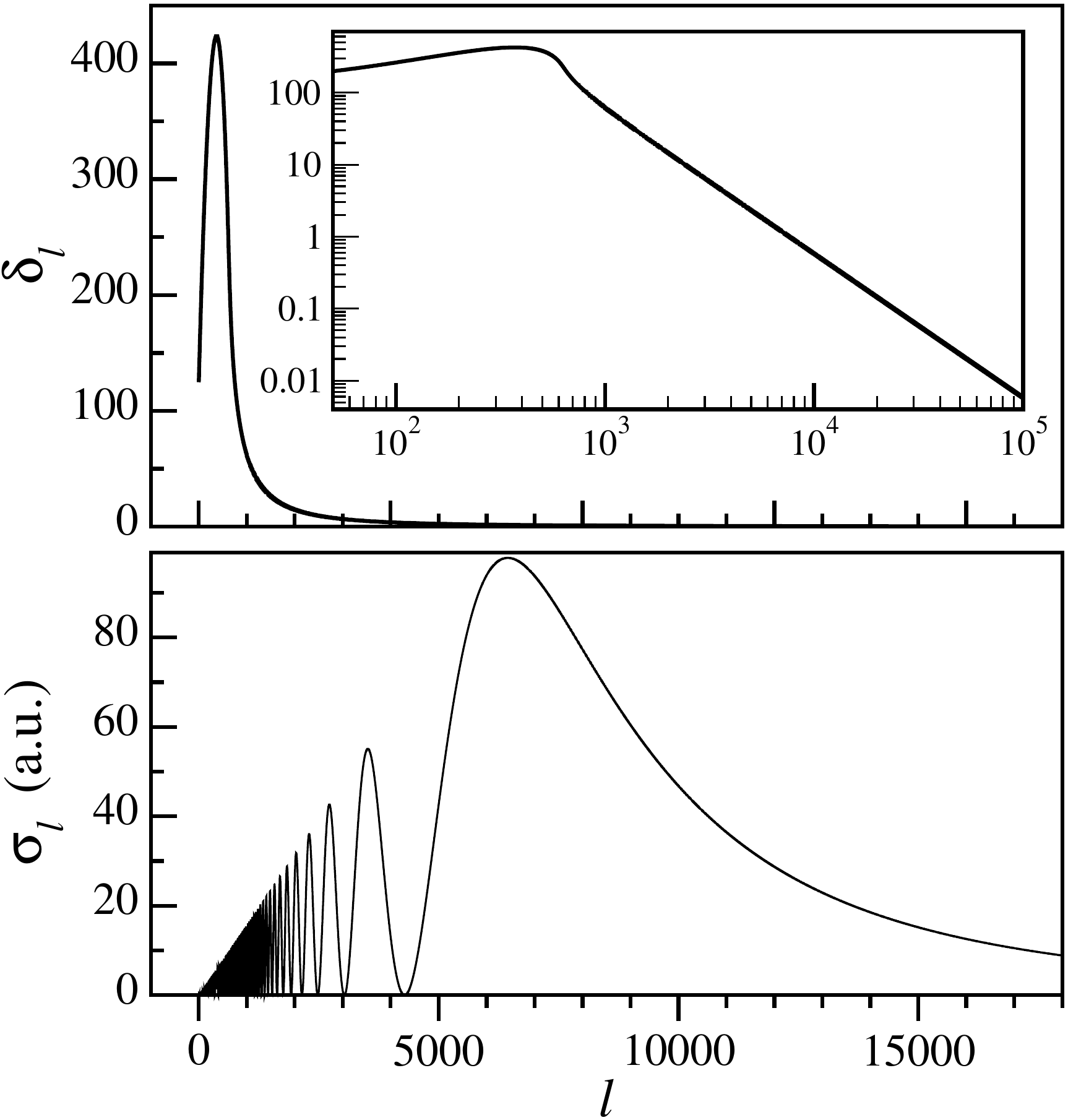}
\caption{\label{fig:delta-vs-ell} Phase shifts and cross section terms
  $\sigma_\ell=\frac{4\pi}{k^2}(2\ell+1)\sin^2\delta_\ell$ for the
  potential energy in Eq.~(\ref{eq:vpotC3}) for $E=0.01$~a.u.  }
\end{figure}

For our second example we use both integral representations, i.e.,
Eq.~(\ref{eq:Delta=int}) for the full phase shift and the two-envelope
formula~(\ref{eq:delta=theta-hat-0}) which yields the phase shift
directly.  We employ the potential energy
\begin{equation}
\label{eq:vpotC3}
V(R) = C_{\rm wall}\exp\left(-\frac R{R_{\rm wall}}\right) 
\;-\; \frac{C_3}{R^3+R_{\rm core}^3},
\end{equation}
with $C_{\rm wall}=10$, $R_{\rm wall}=1$, $R_{\rm core}=5$ and
$C_3=18$ (all in atomic units), and the reduced mass $\mu=\frac m 2$,
where $m$ is the mass of\ $^{88}$Sr.  We computed phase shifts for
$E=0.01\ {\rm a.u.}\approx0.272$~eV for a wide range of partial waves.
The upper panel in Fig.~\ref{fig:delta-vs-ell} depicts the
$\ell$~dependence of the phase shift, while the lower panel shows the
partial-wave terms of the elastic cross section,
$\sigma_\ell=\frac{4\pi}{k^2}(2\ell+1)\sin^2\delta_\ell$.  An exceedingly
large number of partial waves contribute to the cross section; note
that the dominant contribution stems from very high partial waves
($\ell>5000$).  For a fully converged value of the elastic cross
section, we have computed phase shifts up to $\ell=10^5$.

\begin{figure}[t]
\includegraphics[width=\linewidth]{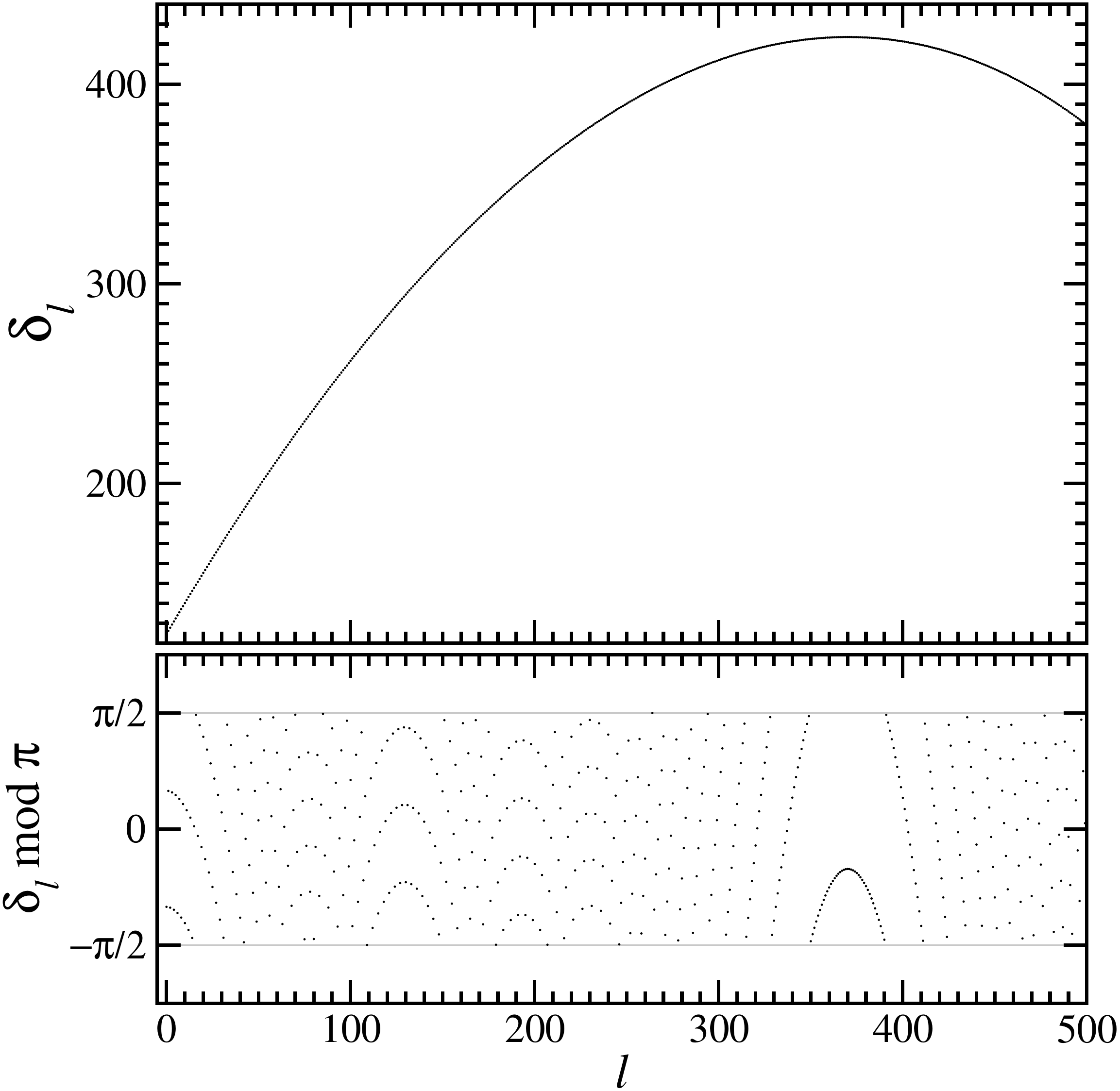}
\caption{\label{fig:mod-pi} Upper: true value of $\delta_\ell$.
Lower:  $\delta_\ell$ mod $\pi$.
Note the vastly different scales used for the vertical axis in the two panels.
}
\end{figure}

Recall that our integral representations yield the true value of the
phase shift (not modulo $\pi$) which has a rather simple
$\ell$-dependence; this suggests that interpolation schemes could be
used to drastically reduce the number of partial waves for which phase
shifts need to be computed.  This added advantage is illustrated in
Fig.~\ref{fig:mod-pi}, where  we compare the true phase shift with
its modulo $\pi$ version.  Moreover, our approach is not restricted to
integer values of $\ell$, and makes it possible to use non-integer
values of $\ell$ as interpolation points; thus, highly accurate
interpolation methods with non-uniform grids, e.g., Chebyshev
interpolation, can be employed.

\begin{figure}[b]
\includegraphics[width=\linewidth]{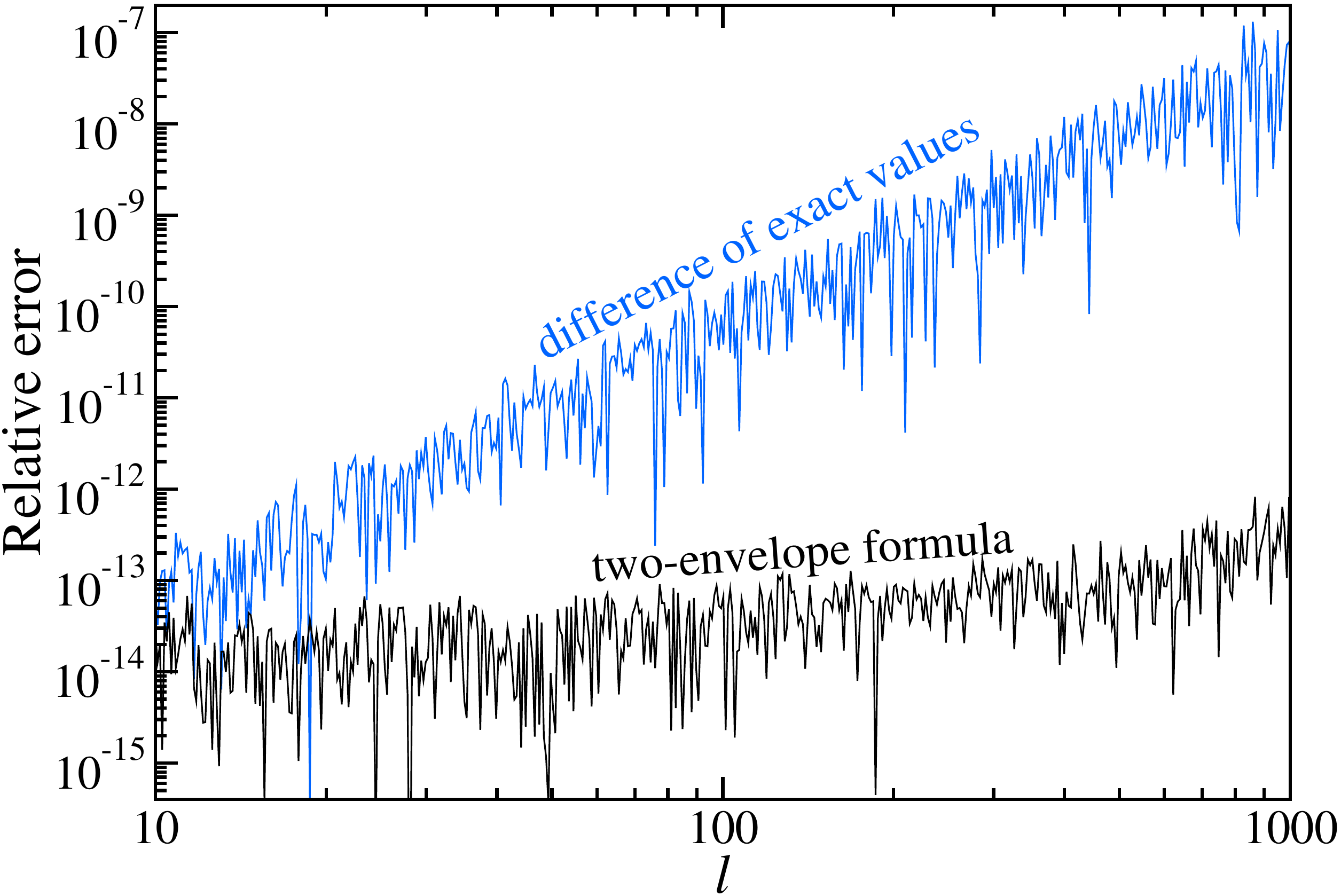}
\caption{\label{fig:noise-bessel} Relative error for the Bessel test.
  Blue line for the difference of the exact Bessel phase shifts, see
  text; black line for the two-envelope formula.}
\end{figure}

Regarding the practical aspects of the computation, some remarks are
in order.  We first emphasize that, using the computational approach
presented in Sec.~\ref{sec:asy}, the simple integral
representation~(\ref{eq:Delta=int}) and the two-envelope
formula~(\ref{eq:delta=theta-hat-0}) can be implemented numerically
such that they both yield accurate results.  However, when using the
integral representation~(\ref{eq:Delta=int}) for the \emph{full} phase
shift, the value of $\delta_\ell$ obtained from
Eq.~(\ref{eq:Delta=theta-at-zero}) will gradually lose precision at
very high $\ell$.  Indeed, for $\ell\to\infty$ we have
$\delta_\ell\to0$ while $\theta(0)\approx\ell\frac\pi2$.  Therefore,
the simple integral representation~(\ref{eq:Delta=int}) must be
avoided at high $\ell$ because of the catastrophic cancellation in
Eq.~(\ref{eq:Delta=theta-at-zero}), even though $\theta(0)$ can still
be computed accurately.  Consequently, when $\ell$ becomes extremely
large, $\delta_\ell$ should instead be computed using the two-envelope formula
derived in Sec.~\ref{sec:two-rho}.

Although the two-envelope formula is highly accurate for all partial
waves, Eq.~(\ref{eq:Delta=int}) has the advantage of much greater
simplicity and could be used at low $\ell$, provided that it is
sufficiently accurate.  A simple rule of thumb exists for finding the
highest partial wave for which
Eqs.~(\ref{eq:Delta=int},~\ref{eq:Delta=theta-at-zero}) yield accurate
results.  Namely, the two-envelope formula needs to replace the simple
formula only if $\ell$ is high enough for the centrifugal term to
become dominant over $V(R)$.  However, at low $\ell$ the simple
formula is highly accurate, as we show next.

\begin{figure}[b]
\includegraphics[width=0.98\linewidth]{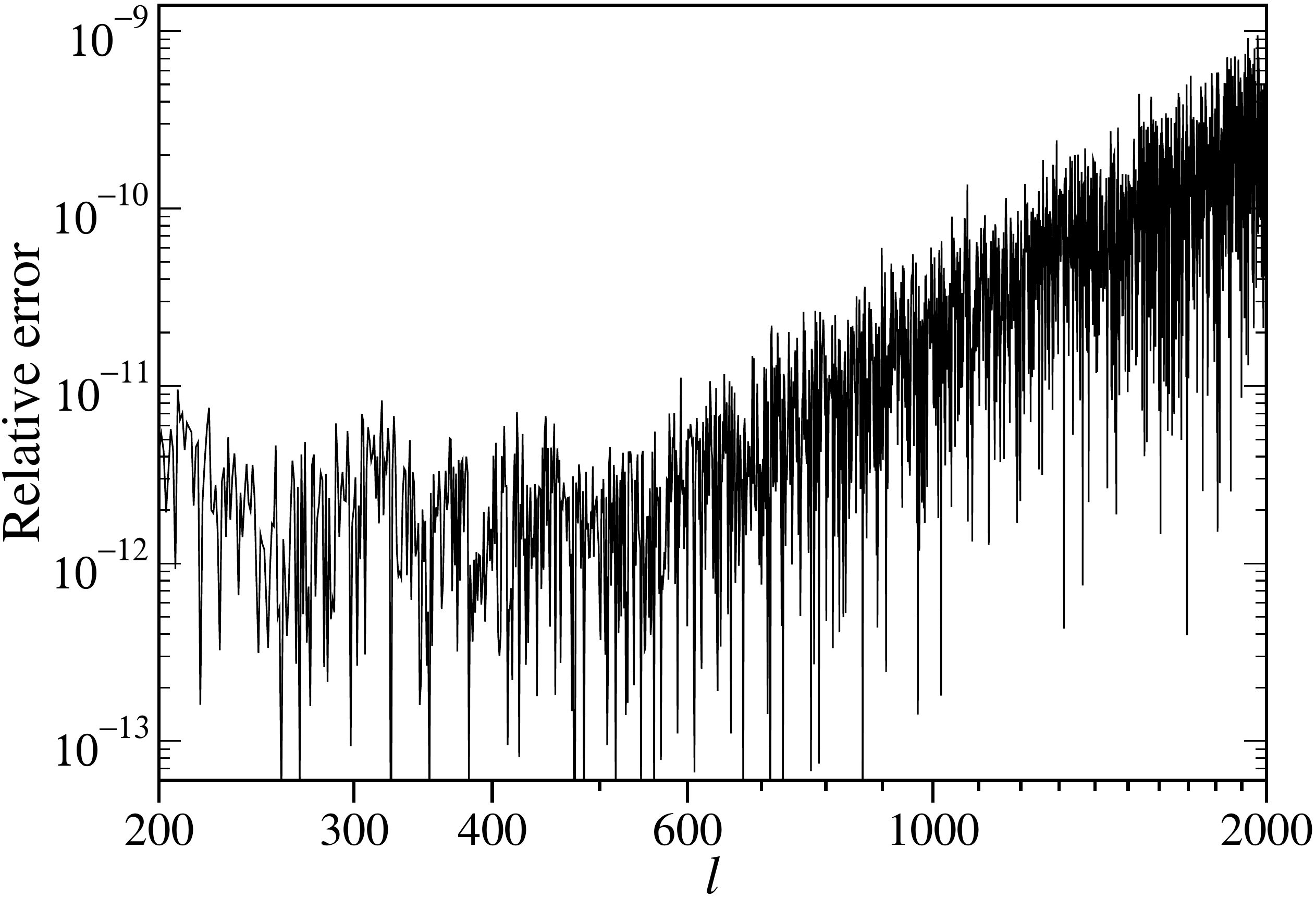}
\caption{\label{fig:noise-compare} Comparison of the simple integral
  representation~(\ref{eq:Delta=int}) and the two-envelope formula
  (\ref{eq:delta=theta-hat-0}).  }
\end{figure}

First, we performed a test for the Bessel case, $V_{\rm eff}
=\frac{\ell(\ell+1)}{2\mu R^2}$, with $V=0$, in order to show that the
two-envelope formula is robust and accurate even for very high $\ell$.
Namely, we computed directly the vanishingly small difference between
the Bessel phase shifts for $\ell_1$ and $\ell_2$ which are nearly
equal.  We used $\ell_1=\ell$ and $\ell_2=\ell-\frac1{\ell^2}$, for
which the exact value of
$\delta_\ell=\delta_B(\ell_2)-\delta_B(\ell_1)$ is
$\delta_\ell=\frac\pi{2\ell^2}$, where $\delta_B(\ell)=-\ell\frac\pi2$
denotes the Bessel phase shift.  The relative error of the numerical
value $\delta_\ell$ obtained with the two-envelope formula is shown in
Fig.~\ref{fig:noise-bessel}, confirming that high accuracy is
preserved at high $\ell$, despite the smallness of $\delta_\ell$.  In
contrast, if the \emph{exact} values of $\delta_B(\ell_{1,2})$ are
subtracted numerically, the loss of accuracy is significant and
becomes catastrophic at very high $\ell$, see
Fig.~\ref{fig:noise-bessel}.  This also illustrates that the failure
of Eq.~(\ref{eq:Delta=theta-at-zero}) at high $\ell$ cannot be
avoided, as it is due to the cancellation of nearly equal quantities.
Nevertheless, Eq.~(\ref{eq:Delta=theta-at-zero}) is sufficiently
accurate at low $\ell$.  Indeed, we performed a test for the simple
integral representation~(\ref{eq:Delta=int}).  Assuming the
two-envelope formula is numerically exact,  the relative error for
$\delta_\ell$ computed using
Eqs.~(\ref{eq:Delta=int},~\ref{eq:Delta=theta-at-zero}) for the
nontrivial potential energy~(\ref{eq:vpotC3}) is shown in
Fig.~\ref{fig:noise-compare}, which makes it readily apparent that the
simple formula is highly accurate for partial waves $\ell\lessapprox
600$, while significant loss of precision only occurs for much higher
partial waves.

\section{Conclusions and future work}
\label{sec:concl}

In this work we presented a novel integral representation for
scattering phase shifts.  A simple version, Eq.~(\ref{eq:Delta=int}),
yields the so called full phase shift, while the two-envelope formula
can be used to obtain $\delta_\ell$ directly.  We emphasize that,
unlike standard approaches which only yield $\delta_\ell$
mod~$\pi$, our integral representations give the true value of the
phase shift.  The integral representations are very useful in
numerical applications, as shown in Sec.~\ref{sec:results}; indeed,
they can be easily implemented to obtain highly accurate results.
Moreover, our approach is based on the phase--amplitude method, which
avoids the explicit computation of highly oscillatory wave functions;
thus, our numerical implementations are very efficient, and are
consistently accurate even for very high partial waves.

We remark that our integral representations are always valid, but the
computational advantages mentioned above rely on the assumption that
the envelope is globally smooth.  However, a globally smooth solution
of the envelope equation~(\ref{eq:rho-linear}) does not always exist;
indeed, as is well known \cite{Multiple_well_Soo_Yin}, when a barrier
separates two classically allowed regions, the solution which is
smooth on one side of the barrier will oscillate on the other side.
To overcome this computational difficulty, it is necessary to develop
an optimization procedure for finding the locally smooth envelopes.
In addition, one needs to find a way to combine the different
solutions which are locally optimized in each classically allowed
region.  This is a nontrivial task, as it requires certain functional
relationships be established between the envelopes and phases
corresponding to disjoint regions which are separated by a barrier.
If such an approach could be rendered computationally feasible, it
would yield the true phase-shift even in the absence of a globally
smooth envelope.

\begin{acknowledgments}

This work was partially supported by the National Science
Foundation Grant No.~PHY-1415560 (DS) and the MURI US Army Research
Office Grant No.~W911NF-14-1-0378 (IS and RC).

\end{acknowledgments}

     \appendix      

     \section{The phase--amplitude approach}

      \label{app:milne}

A brief overview of Milne's phase--amplitude approach is presented here.
According to Milne~\cite{milne}, the \emph{general} solution $\psi$ of
the radial equation (\ref{eq:Schroedinger}) can be expressed in terms
of an amplitude $y$ and a phase $\theta$,
\begin{equation}\label{eq:milne-psi-app}
\psi(R) = c\, y(R) \sin[\theta(R)+\theta_0],
\end{equation}
where $c$ and $\theta_0$ are arbitrary constants.  The amplitude satisfies
the nonlinear equation
\begin{equation}\label{eq:y-milne}
y'' = Uy + \frac{q^2}{y^3},
\end{equation}
with $U=2\mu(V_{\rm eff}-E)$, and the phase $\theta(R)$ is
constructed by integrating
\begin{equation}\label{eq:theta-prime-y}
\theta'=\frac q {y^2}.
\end{equation}
In the equations above, $q$ is arbitrary; the only restriction is
$q^2>0$ in the amplitude equation~(\ref{eq:y-milne}).  We emphasize
that the amplitude and phase appearing in Milne's
parametrization~(\ref{eq:milne-psi-app}) are not unique; indeed, any
solution $y$ of Eq.~(\ref{eq:y-milne}) together with the associated
phase $\theta$ will give a valid representation of $\psi$.  This
undermines the advantage of Milne's method in numerical applications,
because the general solution $y(R)$ of Milne's nonlinear equation has
an oscillatory behavior in classically allowed regions, and the unique
smooth amplitude is very difficult to find
\cite{robicheaux,sidky_phys_essay,matzkin-smooth}.  Despite this
difficulty, Milne's nonlinear equation~(\ref{eq:y-milne}) has long
been used for computational work.  We remark that an \emph{equivalent}
formulation based on a \emph{linear} equation exists 
\cite{discrete_Pinney,leach-2000}, but it remained overlooked in the
physics community until
recently~\cite{jap-rho-linear}.  As we show in
App.~\ref{app:rho-lin-eq}, the linear equation can be obtained by
simply replacing the amplitude with the \emph{envelope} function
$\rho$,
\begin{equation}\label{eq:rho=yy}
\rho(R)=y^2(R).
\end{equation}
The envelope obeys a third order linear differential equation,
\begin{equation}  \label{eq:rho-app}
\rho''' = 4U\rho' + 2 U'\rho.
\end{equation}
In App.~\ref{app:rho-lin-eq} we present two
different derivations for the envelope equation~(\ref{eq:rho-app}),
and in App.~\ref{app:q-invariant} we discuss its equivalence with
Milne's nonlinear equation~(\ref{eq:y-milne}).

Although Eq.~(\ref{eq:rho-app}) is of third order, its linearity makes
it much more convenient than Milne's nonlinear
equation~(\ref{eq:y-milne}).  However, finding the non-oscillatory
solution is still a difficult task.  To overcome this obstacle, we
devised a computational strategy for scattering problems ($E>0$) which
yields the smooth envelope in the asymptotic region, see
Sec.~\ref{sec:asy}.

\section{ The envelope equation}

\label{app:rho-lin-eq}

We present here two different derivations of the envelope
equation~(\ref{eq:rho-linear},~\ref{eq:rho-app}).  The first
derivation is very brief; namely, we substitute $y=\sqrt{\rho}$ in
equation~(\ref{eq:y-milne}) and we find
\[
\rho'' =  2U\rho + \frac1\rho\left[\textstyle \frac 1 2 (\rho')^2 + 2q^2\right].
\]
Next, we multiply both sides  by $\rho$ to obtain
\begin{equation}\label{eq:rho-constraint}
\rho\rho'' = 2U\rho^2 + \frac 1 2 (\rho')^2 + 2q^2,
\end{equation}
which is still a nonlinear equation.  However, we now take the
derivatives of both sides,
\[
\rho\rho''' = 4U\rho\rho' + 2U'\rho^2,
\]
and we divide by $\rho$  to finally obtain Eq.~(\ref{eq:rho-app}).

The second approach is similar to Milne's derivation~\cite{milne} of
Eq.~(\ref{eq:y-milne}).  Namely, we consider two solutions ($\phi$ and
$\chi$) of the radial Schr\"odinger equation~(\ref{eq:Schroedinger}),
and we try to find a differential equation for their product
\[
p \equiv \phi\chi.
\]
Making use of Eq.~(\ref{eq:Schroedinger}) for $\phi$ and $\chi$, we obtain
\[
p'' = 2Up + 2\phi'\chi',
\]
and we now evaluate its derivative,
\[
p''' = 2U'p + 2Up' +2U(\phi\chi'+\phi'\chi),
\]
where we recognize $p'=\phi\chi'+\phi'\chi$, and we find again the
envelope equation~(\ref{eq:rho-app}),
\[
p''' =  4Up' + 2U'p.
\]
As the product of \emph{any} two solutions of the radial
equation obeys the envelope equation, $\phi^2$ and $\chi^2$ are also
valid solutions of Eq.~(\ref{eq:rho-app}), as well as any linear
combination of $\phi^2$, $\chi^2$ and $\phi\chi$
\cite{Pinney_ABC,Chris_Greene_QDT_surface}.
In particular, $\rho=\phi^2+\chi^2$ is  a
valid solution, which corresponds to Milne's ansatz $y=\sqrt{\phi^2+\chi^2}$.

\section{Milne's amplitude equation as a constraint for the envelope equation}

\label{app:q-invariant}

The parameter $q$ appears explicitly in Milne's nonlinear
equation~(\ref{eq:y-milne}).  However, $q$ is absent from the envelope
equation~(\ref{eq:rho-app}), even though it is used when integrating
Eq.~(\ref{eq:theta-prime}) to obtain the phase $\theta$.  This creates
some ambiguity, which stems from the fact Eq.~(\ref{eq:y-milne}) is a
second order differential equation, while the
Eq.~(\ref{eq:rho-linear}) is of third order.  We now try to dispel the
ambiguity and show that the two equations are equivalent.  We first
remark that although $q$ does not appear in Eq.~(\ref{eq:rho-linear}),
it should be assumed implicitly; indeed, if we recast
Eq.~(\ref{eq:rho-constraint}) in the form
\begin{equation}
\label{eq:rho-invariant}
\frac 1 2\rho\rho'' - U\rho^2 - \frac 1 4 (\rho')^2  =  q^2,
\end{equation}
the expression on the left hand side can be interpreted as an
invariant of the envelope equation~(\ref{eq:rho-app}), and any
solution $\rho$ will also obey Eq.~(\ref{eq:rho-invariant}) with a
particular value of $q^2$.  Recall that the equation above is
equivalent with Milne's equation~(\ref{eq:y-milne}), which can thus be
regarded as a constraint for the envelope equation.  Indeed, as we
discuss below, Eq.~(\ref{eq:rho-invariant}) should be used to enforce
the correct initial conditions for $\rho$, such that they correspond to
a fixed value for $q$.

To fully clarify the equivalence of Eqs.~(\ref{eq:y-milne}) and
(\ref{eq:rho-app}), let us compare the sets of initial conditions
required in each case.  When we initialize $\rho$ at $R=R_0$, we
consider given
\[
\rho(R_0), \qquad \rho'(R_0), \qquad \rho''(R_0),
\]
which can be used in the constraint equation~(\ref{eq:rho-invariant})
evaluated at $R=R_0$ to obtain the value of $q$.  Conversely, if $q$
is considered given, we have
\[
\rho(R_0), \qquad \rho'(R_0), \qquad  q = \rm fixed,
\]
which we commonly employ in practice.
Equation~(\ref{eq:rho-invariant}) is now used to obtain $\rho''(R_0)$,
and thus initialize the solution of the envelope equation.
Equivalently, for Milne's amplitude equation we consider given
\[
y(R_0), \qquad y'(R_0), \qquad q = \rm fixed.
\]
Moreover, the families of solutions for different values of $q$ are
all equivalent.  Indeed, if $y_1$ is a solution of
Eq.~(\ref{eq:y-milne}) for a given parameter $q_1$, then
$y_2=(q_2/q_1)^{\frac 1 2}y_1$ is a solution for $q_2$.  Similarly, we
have $\rho_2/q_2=\rho_1/q_1$, and $\theta'_2=\theta'_1$.  Varying the
parameter $q$ is entirely redundant, as the phase $\theta$ remains
unchanged, therefore justifying the convenient choice $q=k$ used
throughout this article.

Finally, we make use of the constraint in Eq.~(\ref{eq:rho-invariant})
to show that the choice $q=k$ is consistent with the initial condition
$\rho=1$ at $R=\infty$, which gives a convenient normalization for the
envelope.  Indeed, when $R\to\infty$, we have $V_{\rm eff}(R)\to0$ and
thus $U(R) \approx U(\infty) = -k^2$ and $U'\approx 0$, while for the
envelope we have
\[
\rho(R)  \approx \rho(\infty), \qquad
\rho'(R) \approx 0,  \qquad
\rho''(R)\approx 0.
\]
Substituting  these asymptotic values in Eq.~(\ref{eq:rho-invariant})
we obtain
\[
-U(\infty)\rho^2(\infty) = q^2,
\]
and using $U(\infty)=-k^2$, we find the parameter $q$,
\[
 q=k\rho(\infty).
\]
Conversely, if one prefers to choose a certain value for $q$, the
equation above yields $\rho(\infty)=q/k$.  However, as shown above,
the normalization constant $\rho(\infty)$
is irrelevant; indeed, we have
\[
\theta'(R)=\frac q {\rho(R)} = k\frac{\rho(\infty)}{\rho(R)},
\]
which ensures
\[
\theta'(R) \approx k, \qquad {\rm when} \ R\to\infty,
\]
and thus the  phase function suitable for scattering problems, as
defined in Sec.~\ref{sec:delta}, has the desired  behavior:
\[
\theta(R) \approx kR,  \qquad {\rm when} \ R\to\infty.
\]


\bibliography{delta_integral}


        \end{document}